\documentclass[9pt,twocolumn,twoside]{pnas-new}

\templatetype{pnasresearcharticle} 

\usepackage{siunitx}
\usepackage{xcolor}

\newcommand{\RH}{$RH$}
\newcommand{\RHamb}{$RH_{\textit{amb}}$}
\newcommand{\Tamb}{$\theta_{\textit{amb}}$}
\newcommand\Rey{\mbox{\textit{Re}}}

\title{Growth of respiratory droplets in cold and humid air}

\author[a,1]{Chong Shen Ng}
\author[a,1]{Kai Leong Chong} 
\author[a]{Rui Yang}
\author[a]{Mogeng Li}
\author[a,b,c]{Roberto Verzicco}
\author[a,d,2]{Detlef Lohse}

\affil[a]{Physics of Fluids Group, Max Planck Center for Complex Fluid Dynamics, J.\,M.\,Burgers Center for Fluid Dynamics and MESA+ Research Institute, Department of Science and Technology, University of Twente, 7500AE Enschede, The Netherlands}
\affil[b]{Dipartimento di Ingegneria Industriale, University of Rome `Tor Vergata', Roma 00133, Italy}
\affil[c]{Gran Sasso Science Institute - Viale F. Crispi, 7 67100 L'Aquila, Italy}
\affil[d]{Max Planck Institute for Dynamics and Self-Organisation, 37077 G{\"o}ttingen, Germany}

\leadauthor{Ng} 

\significancestatement{
Influence of environmental conditions on airborne diseases transmission is an important issue, especially during the pandemic of COVID-19. Human-to-human transmission is mediated by the transport of virus-laden respiratory droplets. Here we investigate the problem from a fluid mechanics perspective by conducting numerical simulations to quantify the fate of respiratory droplets in a warm humid coughing puff under different ambient conditions. We reveal a non-intuitive regime with considerable growth of respiratory droplets, dominated by a super-saturated vapour field, preferentially occurring in cold and humid environments. We further propose a theoretical model that accurately predicts the condition for droplet growth. Our work should inform socializing policies and ventilation strategies for controlling indoor ambient conditions to mitigate dispersion of droplets from asymptomatic individuals.
}

\authorcontributions{C.S.N., K.L.C., R.V., and D.L. designed research; C.S.N., K.L.C., R.Y., M.L., R.V., and D.L. performed research; C.S.N., K.L.C., R.Y., and M.L. analyzed data; C.S.N., K.L.C., R.Y., M.L., R.V., and D.L. wrote the paper.}

\authordeclaration{The authors declare no competing interest.}
\equalauthors{\textsuperscript{1}C.S.N. and K.L.C. contributed equally to this work.}
\correspondingauthor{\textsuperscript{2}To whom correspondence should be addressed. E-mail: d.lohse@utwente.nl}


\begin{abstract}
The ambient conditions surrounding liquid droplets determine their growth or shrinkage. However, the precise fate of a liquid droplet expelled from a respiratory puff as dictated by its surroundings and the puff itself has not yet been fully quantified. From the view of airborne disease transmission, such as SARS-CoV-2, knowledge of such dependencies are critical. Here we employ direct numerical simulations (DNS) of a turbulent respiratory vapour puff and account for the mass and temperature exchange with respiratory droplets and aerosols. In particular, we investigate how droplets respond to different ambient temperatures and relative humidity (RH) by tracking their Lagrangian statistics. We reveal and quantify that in cold and humid environments, as there the respiratory puff is supersaturated, expelled droplets can first experience significant growth, and only later followed by shrinkage, in contrast to the monotonic shrinkage of droplets as expected from the classical view by William F. Wells (1934). Indeed, cold and humid environments diminish the ability of air to hold water vapour, thus causing the respiratory vapour puff to super-saturate. Consequently, the super-saturated vapour field drives the growth of droplets that are caught and transported within the humid puff. To analytically predict the likelihood for droplet growth, we propose a model for the axial RH based on the assumption of a quasi-stationary jet. Our model correctly predicts super-saturated RH conditions and is in good quantitative agreement with our DNS. Our results culminate in a temperature-RH map that can be employed as an indicator for droplet growth or shrinkage.
\end{abstract}

\dates{This manuscript was compiled on \today}
\doi{\url{www.pnas.org/cgi/doi/10.1073/pnas.XXXXXXXXXX}}

\begin{document}

\maketitle
\thispagestyle{firststyle}
\ifthenelse{\boolean{shortarticle}}{\ifthenelse{\boolean{singlecolumn}}{\abscontentformatted}{\abscontent}}{}

The flow physics of respiratory droplet is crucial to understand the airborne transmission of respiratory diseases \cite{bou2021-arfm,mittal2020}, and recently due to the pandemic of COVID-19, the interest on this subject has been renewed \cite{bourouiba2020,asadi2020,setti2020,anfinrud2020,zhang2020,prather2020,jayaweera2020,yang2020towards,bhagat2020effects,abkarian2020stretching}. Tiny saliva and mucus droplets, mostly of the size of tens to hundreds of micrometers \cite{duguid1946,somsen2020}, are expelled from the mouth at speaking, screaming, shouting, singing, coughing, sneezing, or even breathing. The expelled micro-sized droplets have long been recognized to be the carriers for viruses which are responsible for the viral transmission from one host to others \cite{stadnytskyi2020}.

Airborne disease transmission is a difficult subject because many intricate factors simultaneously affect the transmission and a complete understanding requires multi-disciplinary collaboration \cite{moradian2020urgent,ramp2020}. In addition to the understanding from epidemiology \cite{lowen2007influenza,shaman2009absolute,Mecenas2020} and virology \cite{wolfel2020virological}, fluid mechanics also offers essential insight into the mechanisms of airborne disease transmission \cite{mittal2020}. However, surprisingly little is known on the transport and fate of the respiratory droplets, once they have been expelled from the mouth. For example, how the air temperature and the relative humidity influences the actual fate of the droplets remains an open question. It is a complicated process involving interactions between turbulent eddies, vapour field, and temperature field with the respiratory droplets. Therefore, accurate quantification of these effects is necessary. 


On the fate of respiratory droplets in fluid flow, classical work was done by William F. Wells in 1930s \cite{wells1934,wells1936}. The picture that Wells developed, at that time in connection with the transmission of tuberculosis, considers only the settling and evaporation of the droplets. After droplets are expelled from the mouth, they settle to the ground by gravity, and, in this simple model, they evaporate with their surface area decreasing linearly with time, the so-called $d^2$-law \cite{langmuir1918}. Then for a given ambient temperature and relative humidity, one can estimate the lifetime of the respiratory droplet based on the $d^2$-law.

However, in an actual respiratory event, the assumptions of settling and evaporation are insufficient to describe the full physical process. In reality, more complicated flows are possible: Abkarian \textit{et al.}\,\cite{abkarian2020} demonstrated that speech can produce jet-like flow transport, and Bhagat \textit{et al.}\,\cite{bhagat2020effects} illustrated that body or breathing plumes from a person can also spread droplet nuclei in enclosed spaces. Recently, a new paradigm has been suggested by Bourouiba \textit{et al.}\,\cite{bourouiba2014,bourouiba2020,bou2021-arfm}, which states that instead of the isolated droplet being considered, one should also take into account the importance of turbulent vapour puff.
Inspired by this paradigm, using direct numerical simulations, Chong \textit{et al.}\,\cite{chong2020extended} showed that droplets have $O$(100) longer lifetimes in a turbulent vapour puff at high ambient relative humidity than that predicted by the $d^2$-law. 
Past numerical studies that consider different ambient conditions typically adopt mean flow modelling approaches \cite{xie2007,liu2017,wang2020transport}. The role of ambient conditions requires large-scale parameterizations, \cite{balachandar2020host,mittal2020mathematical,chaudhuri2020} but run the risk of obscuring the underlying physics.

In this study, we aim at quantifying the effect of ambient temperature (\Tamb) and relative humidity (\RHamb) on the respiratory droplets using direct numerical simulations of a turbulent puff. Normal coughing with an actual injection profile has been simulated with 5000 droplets being tracked by the Euler-Lagrangian approach. We reveal a regime in which the respiratory droplets considerably grow before their shrinkage due to evaporation. The droplets growth happens under specific combinations of ambient conditions, preferentially in cold and humid environment. 
To predict the criteria of having droplet growth, we further propose a theoretical model to estimate the local relative humidity experienced by the droplets, which agrees excellently to our simulation results. 
Whilst our model can help to formulate public health guidelines or indoor ventilation strategies in order to mitigate the spreading of respiratory droplets, the infectivity of virus-containing droplets and droplet nuclei, which remain after evaporation still requires further investigation, in particular on the virus dose required for infections and whether dried-out droplet nuclei may also be infectious. This issue remains controversial \cite{stadnytskyi2020} and cannot be answered within a fluid dynamical study.

\section*{Results and Discussion}
\begin{figure*}
\centering
\includegraphics[width=1.0\linewidth]{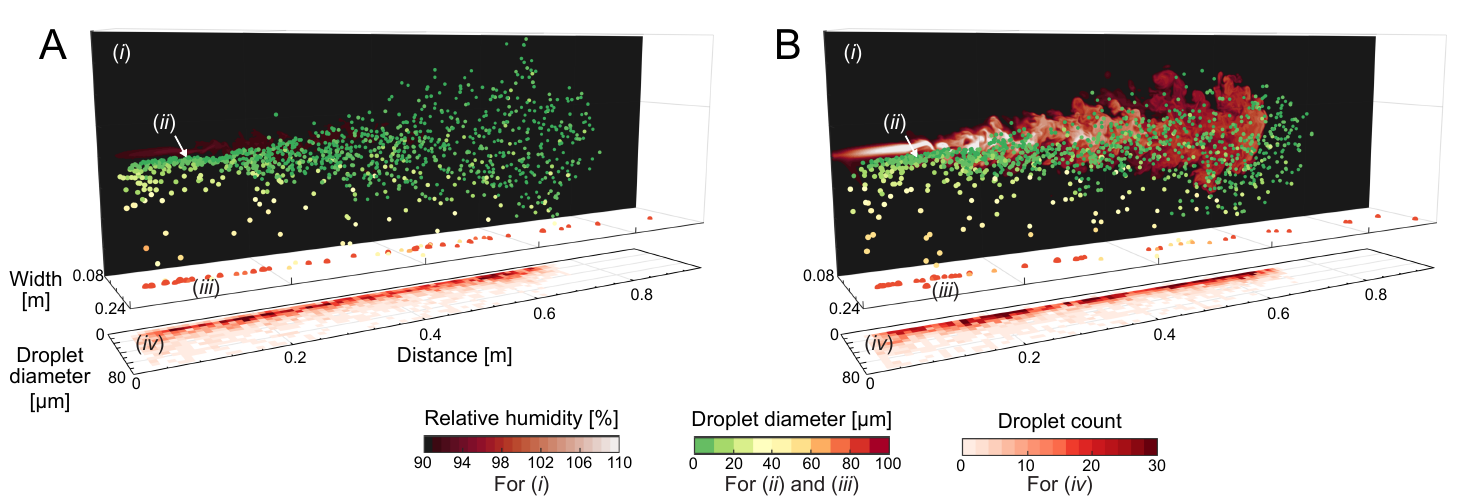}
\caption{Flow visualisation snapshots from our direct numerical simulations of water droplets in a warm humid puff in ambient air at (A) $\theta_{\textit{amb}}=30$\SI{}{\degreeCelsius} and (B) $\theta_{\textit{amb}}=10$\SI{}{\degreeCelsius}, both at \RHamb$~=90$\%. Corresponding movies can be seen in Movies S1 and S2. The snapshots show (i) vertical 2D planes of the local \RH~fields, (ii) the instantaneous droplets spatial distribution, (iii) the heavy large droplets which already fell on the ground, and (iv) the instantaneous droplet size histograms versus distance. 
The local \RH~planes are taken from the vertical mid-plane of the puff and are plotted on the background for clarity. Droplets are colour coded by their instantaneous sizes. 
Initial droplet sizes are prescribed with a distribution similar to J.\,P.\,Duguid \cite{duguid1946} (plotted in Fig.\,S2) and are injected evenly in time with the same local inflow velocity. The initial temperature of the droplets and puff is 34\SI{}{\degreeCelsius}. Both snapshots are taken at 0.6\SI{}{\second} corresponding to the cut-off time of the puff. In the colder conditions of (B), the expelled humid puff over-saturates (seen as the lighter colored \RH~field and visible in Movie S2), which in turn dictates growth of smaller droplets caught within the puff. Correspondingly, the droplet counts are confined within a narrower range of sizes in (B) as compared to (A), see Fig.\,\ref{fig:fig2}A.}
\label{fig:fig1}
\end{figure*}

\begin{figure}[h!]
\centering
\includegraphics[width=0.8\linewidth]{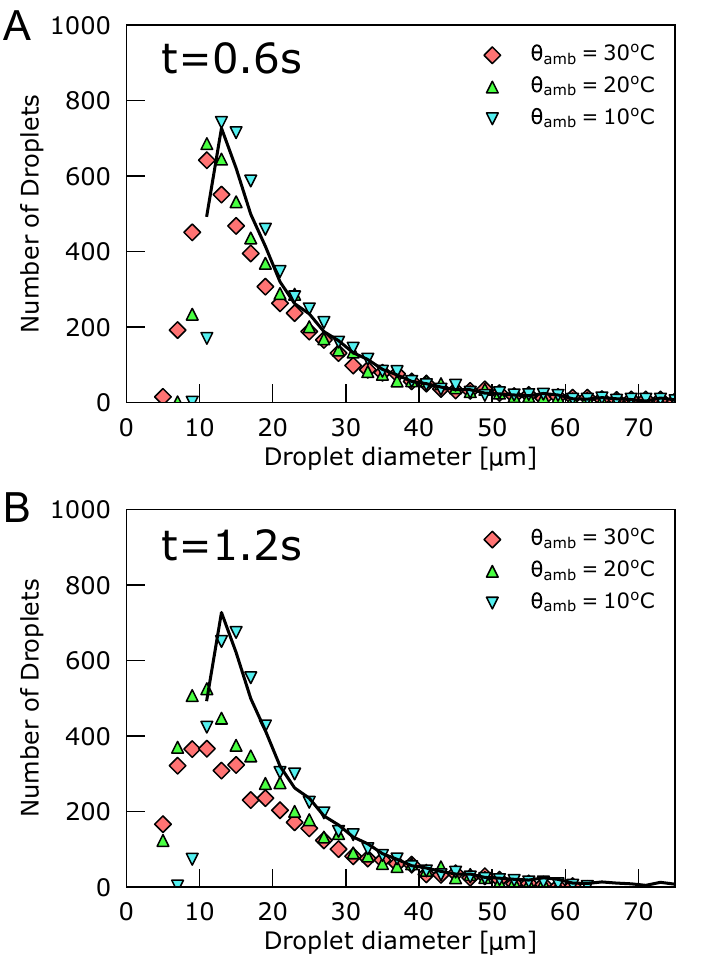}
\caption{Histogram of droplet sizes at (A) $t = 0.6$s and (B) $1.2$s for \Tamb~$=10, 20, 30$\SI{}{\degreeCelsius}. As a point of reference, the solid lines denote the initial droplet size distributions. The bin size of the histogram is $2$\SI{}{\micro\meter}.
At $t=0.6$s (panel A), the peak of the histogram for $\theta_\textit{amb}=30$\SI{}{\degreeCelsius} marginally shifts down, whereas the overall histogram for $\theta_\textit{amb}=10$\SI{}{\degreeCelsius} remains roughly similar to the initial distribution. The gap is more pronounced at $t=1.2$s (panel B): For $\theta_\textit{amb}=30$\SI{}{\degreeCelsius}, the peak of the histogram is halved from the initial peak, whereas in contrast, for $\theta_\textit{amb}=10$\SI{}{\degreeCelsius}, the histogram still remains roughly unchanged. The relative invariance of the histogram for $\theta_\textit{amb}=10$\SI{}{\degreeCelsius} is due to the droplet growth dictated by the super-saturated vapour puff, which is in turn determined by cold and humid ambient conditions, as described in the main text.}
\label{fig:fig2}
\end{figure}

\subsection{Effect of Ambient Temperature on Turbulent Vapour Puff and Droplet Counts}
First, we highlight the differences in the local RH values for two different \Tamb~(10\SI{}{\degreeCelsius} and 30\SI{}{\degreeCelsius}) at \RHamb~$=90$\%. Note that \Tamb~and \RHamb~both constitute the control parameters of our study. The local $RH$ field is shown in figure \ref{fig:fig1} together with the instantaneous droplet distributions and instantaneous count histogram. 
At $\theta_{\textit{amb}}=30$\SI{}{\degreeCelsius}, the local \RH~field is almost indistinguishable from \RHamb, as shown in Fig.\,\ref{fig:fig1}A. In contrast, at $\theta_{\textit{amb}}=10$\SI{}{\degreeCelsius} (Fig.\,\ref{fig:fig1}B), the local \RH~field reaches super-saturation values larger than 100\% (whitish colour shading). 
In the count histograms in Fig.\,\ref{fig:fig1}B, most of the droplet diameters remain at least around \SI{20}{\micro\meter} throughout the spread distance, whereas in Fig.\,\ref{fig:fig1}A, droplet diameters can go below \SI{20}{\micro\meter}. This means that droplets expelled into colder ambient surroundings shrink slower because of the super-saturated local \RH~field. On the other hand, droplets expelled into hotter ambient surroundings shrink much more rapidly because the local RH rapidly under-saturates.

To illustrate the temperature sensitivity of the droplet size distributions, in Fig.\,\ref{fig:fig2}, the distributions across three \Tamb, $10$\SI{}{\degreeCelsius}, $20$\SI{}{\degreeCelsius}, and $30$\SI{}{\degreeCelsius} are shown at $t=0.6$\SI{}{\second} (Fig.\,\ref{fig:fig2}A) and 1.2\SI{}{\second} (Fig.\,\ref{fig:fig2}B). In the figure, the initial droplet size distribution (solid line) is also included for comparison. 
In Fig.\,\ref{fig:fig2}A, which corresponds to the cut-off time of the simulated puff, many smaller droplets $\lesssim 15$um already exist at $\theta_{\textit{amb}}=20$\SI{}{\degreeCelsius} and $30$\SI{}{\degreeCelsius} as compared to $\theta_{\textit{amb}}=10$\SI{}{\degreeCelsius}. 
Interestingly, the histogram distribution of droplet diameters for \Tamb~$=10$\SI{}{\degreeCelsius} is slightly higher and shifted to the right from the initial droplet size histogram, reflecting that a portion of the smaller droplets in fact grew. The reason for this increase is that due to the cold surrounding air, which can take up less vapour than the warm one, the puff gets super-saturated. This effect will be quantified in detail in Section B. 
At the later time of $t=1.2$\SI{}{\second} (Fig.\,\ref{fig:fig2}B), most of the droplet sizes for $\theta_{\textit{amb}}=20$\SI{}{\degreeCelsius} and $30$\SI{}{\degreeCelsius} are shifted to the left towards much smaller sizes. Remarkably, the size distribution for $\theta_{\textit{amb}}=10$\SI{}{\degreeCelsius} remains roughly similar to the initially prescribed size distribution, with some lower counts at sizes $\lesssim 15$\SI{}{\micro\meter}. 

After having shown the distinct droplet shrinkage behaviour under different ambient conditions, in the following section, we will further and in more detail quantify how the local fluid properties affect the shrinkage rate of the droplets. 


\begin{figure}[h!]
\centering
\includegraphics[width=1.0\linewidth]{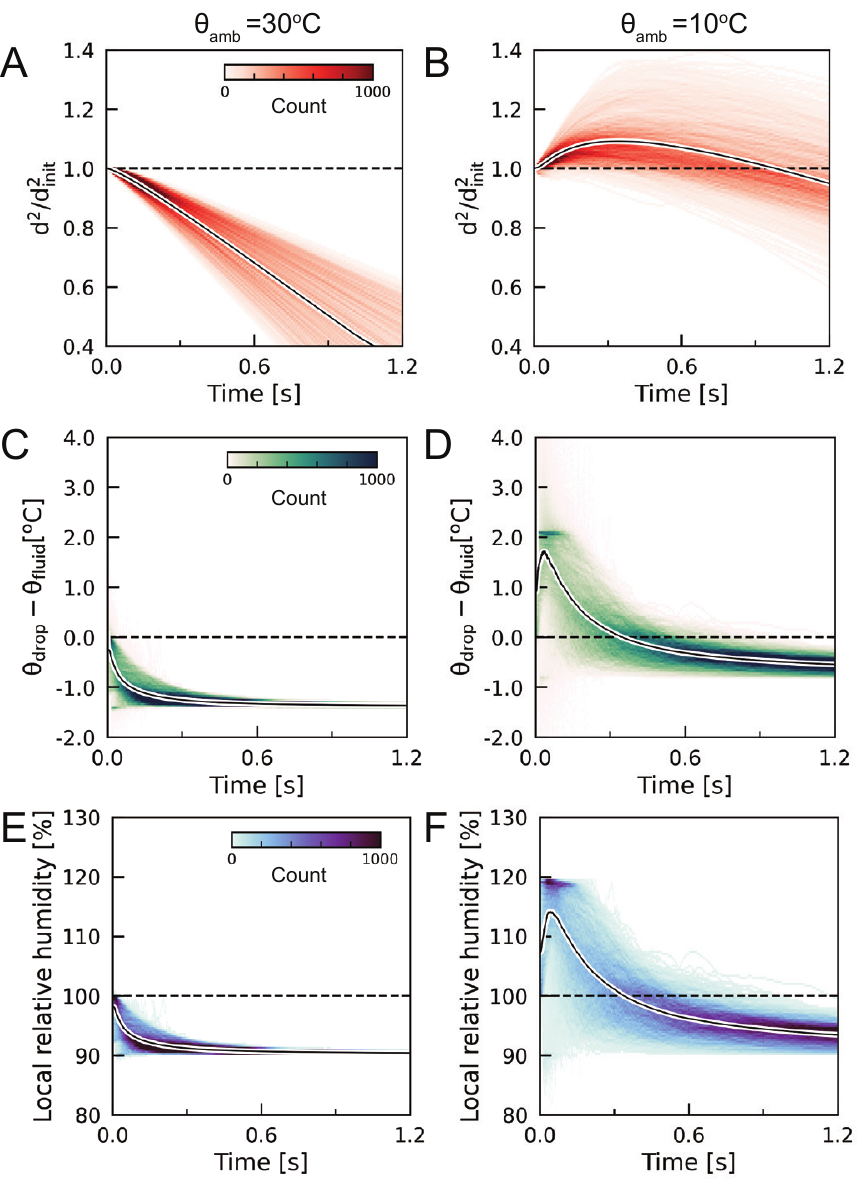}
\caption{\label{fig:timechange} Lagrangian statistics of droplets with initial droplet diameters $d_{\textit{init}} \approx$ \SI{15}{\micro\meter} for $\theta_{\textit{amb}}=30$\SI{}{\degreeCelsius} and $10$\SI{}{\degreeCelsius}. (A,B) Change of droplet surface area $d^2/d^2_{\textit{init}}$ versus time, (C,D) Temperature difference between droplets and surrounding fluid, $\theta_{\textit{drop}}-\theta_{\textit{fluid}}$, versus time, and (E,F) Local \RH~versus time experienced by droplets. The time is shifted to the initial expelled time of each droplet. The ensemble average value is shown by the solid curve. The droplet counts are indicated by the colors, see inset colorbars. For $\theta_{\textit{amb}}=30$\SI{}{\degreeCelsius} (in A), droplets undergo pure evaporation after being expelled. Conversely, for $\theta_{\textit{amb}}=10$\SI{}{\degreeCelsius} (in B), most droplets undergo growth in the initial stages and then evaporate. This growth in (B) is dictated by the super-saturated local \RH~values (see F and visible in Movie S2), which exceed 100\%.}
\label{fig:fig3}
\end{figure}

\subsection{Lagrangian Statistics of the Droplets}
For the following Lagrangian statistics, we focus on the droplets with initial diameter around 15\SI{}{\micro\meter}, given that for those we observe the most pronounced temperature effect, see the count histogram in Fig. \ref{fig:fig2}B. First, we track the droplet surface area $\propto d^2$ versus time as shown in Fig. \ref{fig:fig3}A,B. Note that droplets are expelled from the mouth at different time instants until the respiratory event stops ($0.6$\SI{}{\second} for the coughing event considered here). In order to compare the droplet statistics, we have shifted the time frame by their respective ejecting times. 

For \Tamb~$=30$\SI{}{\degreeCelsius} (Fig.\,\ref{fig:fig3}A), the normalized surface area of the droplets $d^2/d^2_\textit{init}$ decreases monotonically with time (the ensemble average of the value is depicted by the solid curve), which indicates pure evaporation of the expelled droplets. The surface area decreases linearly with time with the averaged droplets surface area being halved after 1\SI{}{\second}. Although a linear decline of the surface area is consistent with the classical $d^2$-law \cite{langmuir1918}, one should note that the magnitude of the droplet shrinkage rate is still much smaller than that predicted by William F. Wells \cite{wells1934}. Such a discrepancy is caused by the fact that the turbulent vapour puff engulfs small droplets and the puff contains higher \RH~than the ambient fluid \cite{bourouiba2020,chong2020extended}.

Surprisingly, for the low \Tamb~of $10$\SI{}{\degreeCelsius} (Fig.\,\ref{fig:fig3}B), the respiratory droplets evolve in a very different and distinct way: Instead of the pure evaporation during the time evolution of the droplets, there exists an initial stage (from the expelled time to about $0.4s$) during which the droplets can grow with the averaged $d^2$ increasing by $10\%$. This non-monotonic trend of the shrinkage markedly differs from the classical $d^2$-law. It also contradicts the well-accepted view that the respiratory droplets should simply evaporate after expulsion. Such a phenomenon of droplets condensation during the respiratory event is the first time, to the best of our knowledge, to be quantified thanks to the Lagrangian statistics obtained from DNS, which has been neglected in the hitherto studies of droplet transmission in exhalation.

Another indicator that for $10$\SI{}{\degreeCelsius} the droplet size first increases is that the droplet temperature $\theta_\textit{drop}$ increases relative to the fluid temperature at the location of the droplet $\theta_\textit{fluid}$ (Fig.\,\ref{fig:fig3}D). As latent heat is released during droplet condensation for $10$\SI{}{\degreeCelsius} (Fig.\,\ref{fig:fig3}D), the difference $\theta_\textit{drop}-\theta_\textit{fluid}$ increases during the early time. In contrast, $\theta_\textit{drop}-\theta_\textit{fluid}$ remains less than zero for \Tamb~$=30$\SI{}{\degreeCelsius} (Fig.\,\ref{fig:fig3}C), due to the latent heat needed for evaporation. Eventually, the droplet temperature saturates at a value lower than the fluid temperature. This is reasonable because the Jakob number, which is the ratio of the sensible heat to the latent heat due to the phase change, is much smaller than one (0.04 and 0.007 for \Tamb~$=10$\SI{}{\degreeCelsius} and $30$\SI{}{\degreeCelsius}, respectively). Therefore, the evaporation cooling effect plays a significant role here.

\begin{figure}[t!]
\centering
\includegraphics[width=0.8\linewidth]{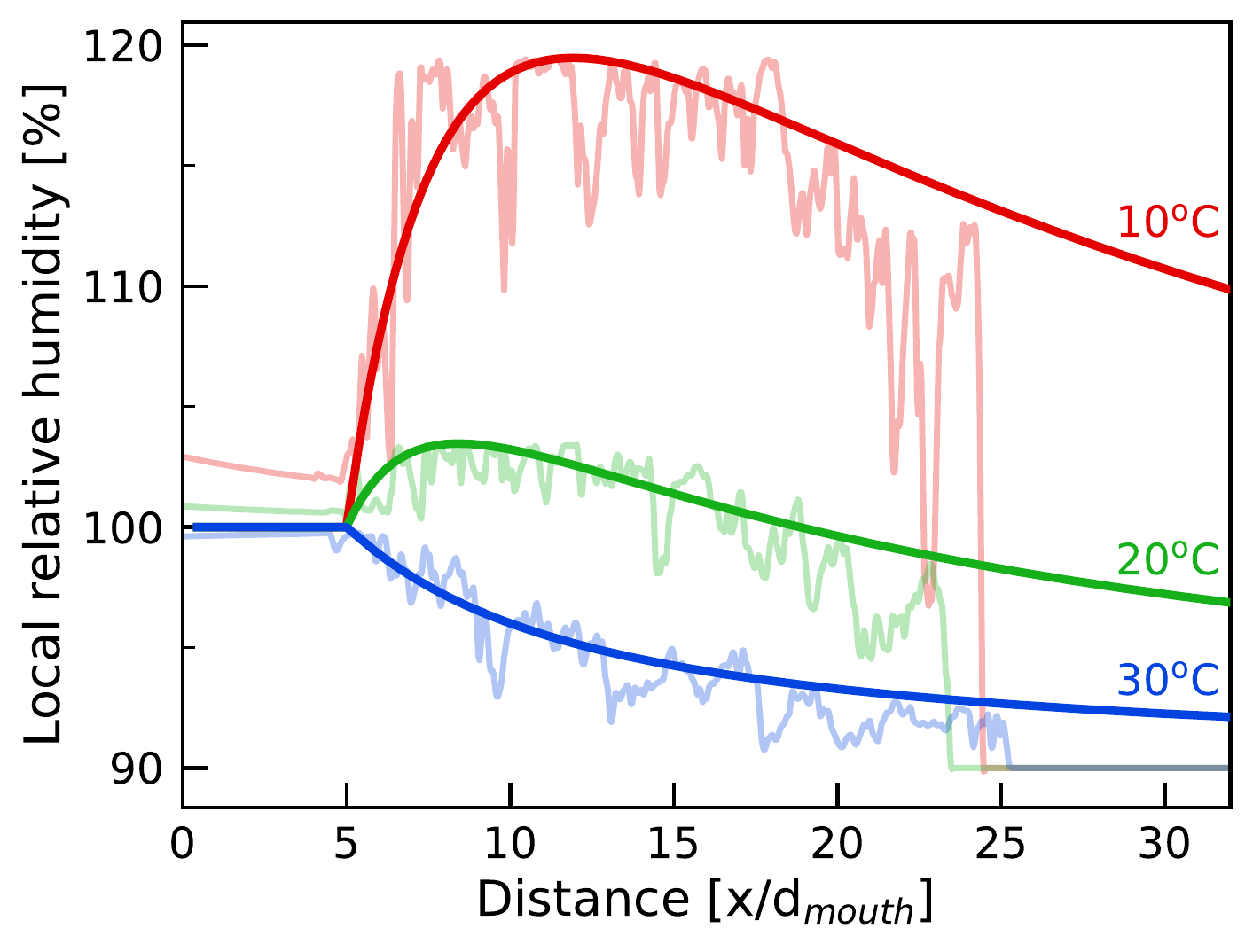}
\caption{Comparison of the results from our present DNS results (light curves) with those from the model for axial \RH~derived from \eqref{eqn:axialscalarmodel} (smoother darker curves) at $t=0.2$s and \RHamb~$=90$\%. The data is plotted versus distance normalized by the mouth diameter, $x/d_{\textit{mouth}}$. The different coloured curves represent data for \Tamb~of 10\SI{}{\degreeCelsius} (red), 20\SI{}{\degreeCelsius} (green), and 30\SI{}{\degreeCelsius} (blue), respectively. The model \RH~curves all extend beyond $x/d_{\textit{mouth}}\approx25$ since the model assumes a fully developed jet, which is not the case for the present simulated case of a puff. However, the near-field axial model accurately captures the puff, implying quasi-stationary self-similar jet characteristics.}
\label{fig:fig4}
\end{figure}

To understand why the droplets can grow during the early time, a key quantity to examine is the local \RH~experienced by the droplet. In Fig.\,\ref{fig:fig3}E,F, we compare the time evolution of the local \RH~at the droplet location for the two different \Tamb, showing quite different behaviours: For $30$\SI{}{\degreeCelsius}, the local \RH~decreases monotonically to \RHamb~which is $90\%$ in this case. However, for $10$\SI{}{\degreeCelsius}, the droplets first experience super-saturation within $0.4$\SI{}{\second} after expelling and the maximum \RH~reaches almost $115\%$. This duration of having super-saturation is consistent with the duration of droplet growth, which reflects that it is the local super-saturation of the surrounding air which makes the droplets grow.

In fact, our daily life experience gives us intuition about this intriguing droplet condensation phenomenon. In cold and humid weather, one may observe the ``white mist'' coming out from the mouth while breathing or speaking. The physical explanation is that: Warm air can contain more moisture than cold air. Therefore, the exhaled warm vapour puff becomes super-saturated when it enters the cold ambient fluid, and the droplet nuclei and dust in air favours vapour condensation. Such super-saturated conditions become more prominent if the temperature difference between the exhaled vapour and ambient fluid is larger. Here, we have clearly demonstrated the significance of super-saturation in the respiratory event that leads to the possibility of vastly different droplet dynamics.

\subsection{Theoretical Model Predicting Relative Humidity Profiles and Super-saturation Criteria}
Motivated by the strong effect of the local \RH~field on the droplets, here, we propose a simple model to calculate the local \RH. Several assumptions first need to be made to justify this model. First, the axial scalar quantities (vapour mass fraction and temperature) are the dominant fluid properties that determine droplet growth/evaporation, and secondly, the exhaled puff at the time of $t=0.2$\SI{}{\second} exhibits jet-like properties such that the quasi-stationary mean properties of the puff admits self-similarity. Here, we employ the Antoine equation in order to compute the saturated vapour mass fraction for a given temperature field \cite{russo2014}. In short, given the local values of vapour mass fraction and temperature along the jet centreline, the axial \RH~can be directly computed.

\begin{figure}[t!]
\centering
\includegraphics[width=0.9\linewidth]{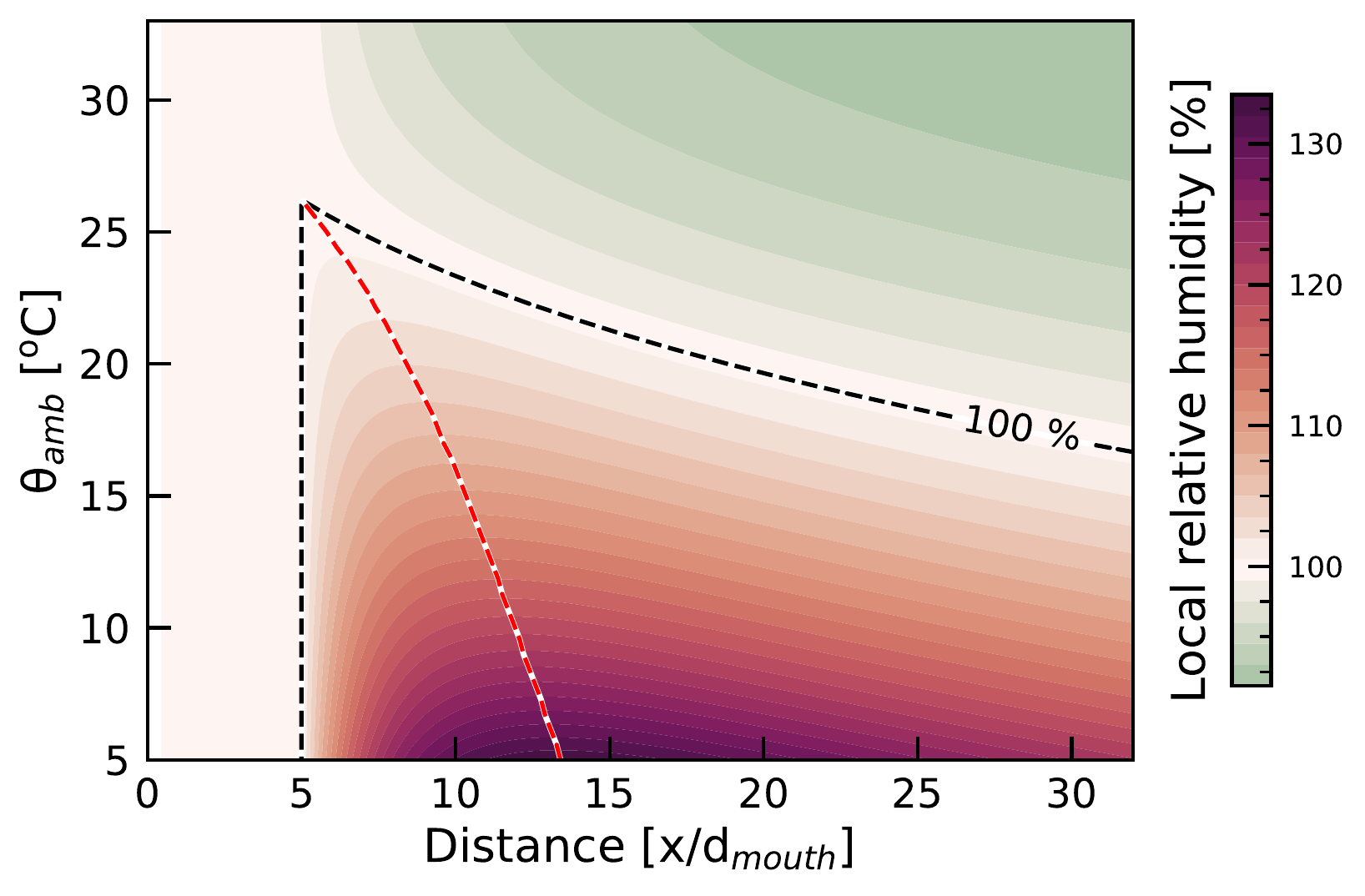}
\caption{\Tamb~versus distance map for the \RH~model at $t=0.2$s. The contours are computed for \RHamb$~= 90$\% with the color coded by the local \RH. The boundary for \RH~oversaturation (local \RH$~> 100$\%) is shown by the dashed line. With decreasing \Tamb, the range with local \RH$~> 100$\% (width of the reddish area under the dashed curve) increases, which promotes growth of the smaller droplets because of the longer exposure to super-saturated regions. The most serious super-saturation conditions in this plot are for $\theta_\textit{amb}=5\SI{}{\degreeCelsius}$ and $x/d_{\textit{mouth}}\approx 10-15$, namely beyond a local \RH~of $130\%$. For a given $\theta_\textit{amb}$, the location of the maximum is marked by a dashed red line.}
\label{fig:fig5}
\end{figure}

Following the theory for axisymmetric turbulent jets \cite{hinze1949transfer,schlichting1968boundary,antoine2001turbulent}, the axial scalar concentration distribution (say $\Phi_\textit{axial}$) can be approximated by the relation,
\begin{equation}
    \frac{\Phi_\textit{axial}}{\Phi_\textit{inlet}} = \frac{5}{1+x/d_\textit{inlet}} \label{eqn:axialscalarmodel}
\end{equation}
where $\Phi_\textit{inlet}$ is the inlet scalar concentration value, and $x/d_\textit{inlet}$ is the axial distance normalised by the inlet diameter. Since this model diverges as $x/d_\textit{inlet} \rightarrow 0$, we further impose the requirement that max[$\Phi_\textit{axial}/\Phi_\textit{inlet}$]$=1$. Both the exhaled vapour mass fraction and temperature are modelled using \eqref{eqn:axialscalarmodel} along the axis normal to the inlet.

To assess the validity of our assumptions, we plot the resulting model of the axial \RH~profiles for different \Tamb~in Fig.\,\ref{fig:fig4}. The agreement with our DNS (lighter curves) is quite promising, which is consistent with our assumption that the puff retains self-similar jet-like characteristics at cut-off time. This agreement lends some confidence to our ability to model the axial \RH~of our flow at a specific time instant, while the cough will approximate to a turbulent puff or an interrupted jet at large $t$ after the injection velocity becomes negligible \cite{wei2017human}.

Based on the robustness of this model, we further propose the use of the maximum local \RH~as a metric to indicate the likelihood of the growth for small droplets. To illustrate this point, in Fig.\,\ref{fig:fig5}, we plot the axial \RH~map for various \Tamb~versus distance at the fixed \RHamb~of 90\%. The dashed line shows the crossover between super-saturated and under-saturated axial \RH~values. One interesting observation from this model is that with decreasing \Tamb, the overall axial distance that experiences super-saturation increases. In other words, it is more likely that with decreasing \Tamb~(at \RHamb~$=90$\%) small droplets will encounter longer super-saturated distances and grow in size. This likelihood to grow is important in understanding how far the droplets can be advected by the puff in the initial stages of the cough, before evaporating and dispersing.

\begin{figure}[th!]
\centering
\includegraphics[width=0.9\linewidth]{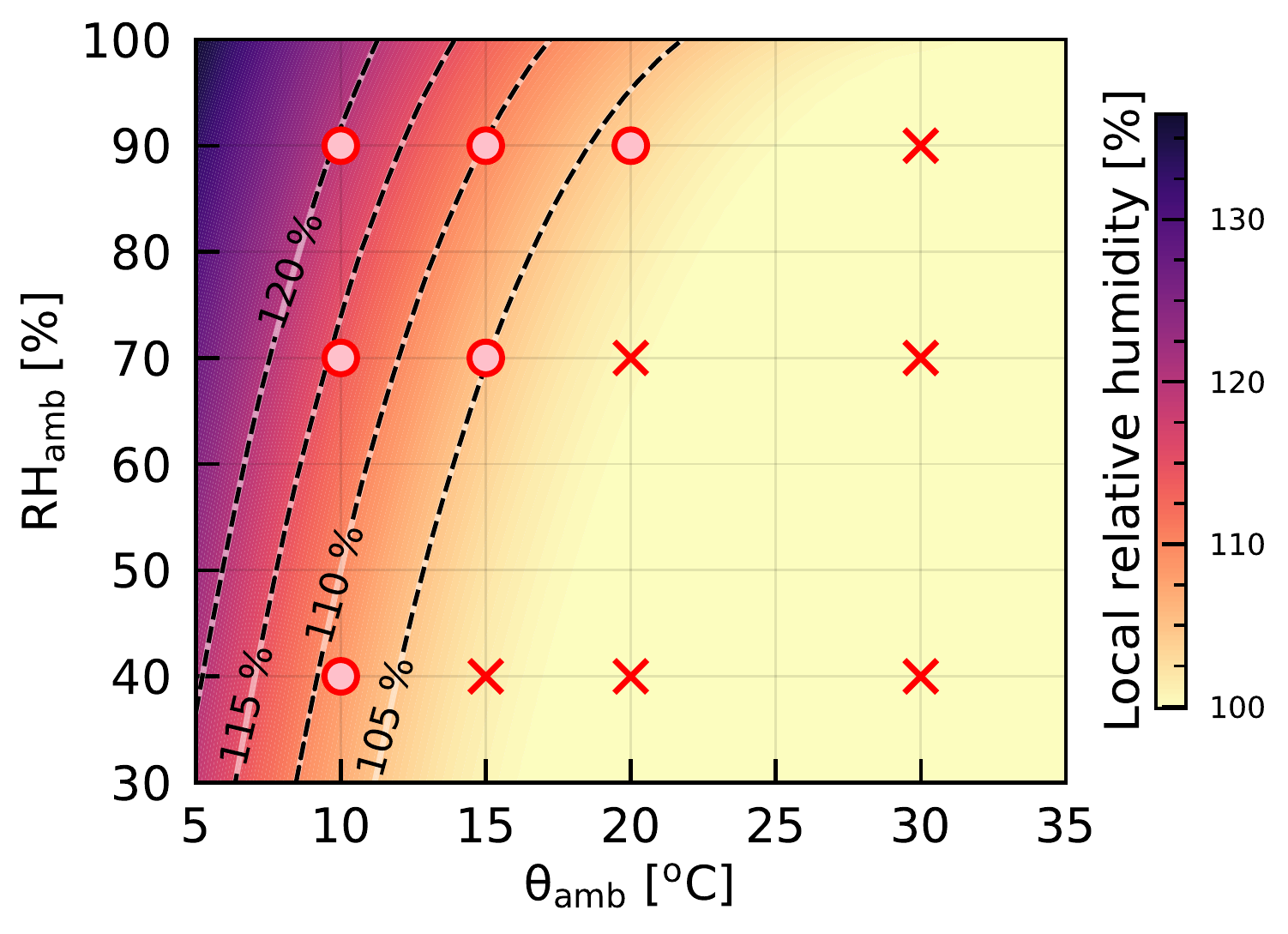}
\caption{Map of maximum local \RH~super-saturation values (see Fig.\,\ref{fig:fig5}) for different \Tamb~and \RHamb~based on the \RH~model. For the case of $RH_\textit{amb}=90\%$ these are the values of the maximal local \RH~in Fig.\,\ref{fig:fig5} (red dashed curve). The values can be used as a gauge to indicate the tendency of droplets to experience super-saturation, given that the droplets are caught within the turbulent puff. The color is coded by the upper bounds for super-saturation. The plot indicates that in cool and humid environments, small droplets within a coughing puff will tend to grow once expelled. To assess the performance of our model, we also plot our DNS results for droplets with an initial size of $\approx 15$\SI{}{\micro\meter} and indicate growth or shrinkage using circle and cross symbols, respectively. (The detailed breakdown is plotted in Fig.\,S1.) A good agreement between our DNS and the \RH~model is seen.}


\label{fig:fig6}
\end{figure}

Finally, we extend our idea of modelling the axial \RH~to map out the maximum local \RH~values for a range of \RHamb~and \Tamb. This map is shown in Fig.\,\ref{fig:fig6}. From the figure, we find that the maximum local \RH~values are dependent on both \RHamb~and \Tamb. 
Several interesting observations can also be made from the figure. For instance, at a given indoor ambient conditions of \RHamb~$=40$-$50$\% (within acceptable occupancy comfort levels), the maximum local \RH~is at most 100\% as long as $\theta_\textit{amb}>20$\SI{}{\degreeCelsius}. However, once $\theta_\textit{amb}<20$\SI{}{\degreeCelsius}, the local \RH~tends to super-saturate (\RH$~>100$\%), implying that in cooler environments, smaller droplets within the humid puff will first undergo supersaturation-driven growth in the early stages of the cough before evaporating upon fall-out. 
Such conditions become particularly exacerbated in outdoor environment, for example, in stadiums or sports field during autumn or winter, when the temperatures tend to be colder. Similar conditions can also occur in crowded bus and train stations.  Super-saturated conditions are more easily attainable (easily seen in exhaled mists) and therefore alters the respiratory droplet dynamics. 

To substantiate our proposed model, in Fig.\,\ref{fig:fig6}, we indicate on the plot 12 different control parameter combinations (\RHamb~and \Tamb) for which droplets with an initial size of $\approx15$\SI{}{\micro\meter} grow (indicated by a circle symbol) or shrink (indicated by a cross symbol) in our DNS and compare with the predictions of the model. (The profiles of $d^2$ versus time is provided in Fig.\,S1 of the Supplementary Material). By comparing the symbols from the DNS and the map from the model, a clear correlation between growth and super-saturated local \RH~fields can be seen. Thus we propose that this map can be used as an indicator to determine the droplet size distributions of a similar cough. 

It is also worth noting that we did not consider the influence of background wind, which is an important aspect both in ventilated indoor and outdoor environments. We also stress that we make no claims to correlate growth or shrinkage of droplets with virus transmissions or infectivity. The subject of the viability of droplet sizes for viral media is still the topic of ongoing investigations in other communities and are not the focus of this work.

\section*{Conclusions}
Through the Lagrangian statistics of the respiratory droplets, we have identified that when droplets are expelled from the mouth to the cold environment with high enough relative humidity, the droplet will first grow instead of immediately shrink, because the turbulent vapour puff becomes super-saturated. More importantly, we have provided a theoretical framework to accurately predict (i) how the relative humidity varies with distance in a respiratory puff, and (ii) the threshold of ambient temperature (\Tamb) and relative humidity (\RHamb) for which the super-saturated vapour field can result. Our theoretical model can also be applied to other respiratory events whenever there is jet-like transport, such as speaking \cite{abkarian2020}, which is extremely relevant for asymptomatic and presymptomatic spreading of the coronavirus. In the situation of continuous speaking, when the hot vapour is continuously injected into the cold environment, we expect that the super-saturation zone may remain in front of the mouth for a long period of time. Our finding implies that tiny droplets of the size of tens of micrometers can propagate further away when the weather is cold and humid due to the protection from the super-saturated vapour field. The result of the very distinct fate of respiratory droplets under different \RHamb~and \Tamb~may be instructive for developing timely strategies in mitigating the COVID-19 pandemic in different seasons. For example, during the Winter time, where the maritime climate for many coastline European countries have much higher \RHamb, the transmission in outdoor environments can presumably not be overlooked. 
However, we note that obviously also non-fluid-dynamical aspects affect the transmission and infectivity of viruses,
such as from virology, epidemiology, and sociology, to name a few. 
These are not covered within our study and presumably interfere with our fluid dynamical results. 

\matmethods{
\subsection*{Governing equations}
We consider an incompressible flow ($\partial {u}_i / \partial {x}_i = 0$) of gas phase, with both temperature and vapour concentrations coupled to the velocity field by employing the Boussinesq approximation. The governing equations read:
\begin{equation}
    \frac{\partial{u_i}}{\partial t} + u_j \frac{\partial{u_i}}{\partial x_j} = -\frac{\partial p}{\partial x_i} + \nu_{air}\frac{\partial^2{u_i}}{\partial x_j^2} + g (\beta_\theta \theta + \beta_c c)\hat{e}_y,
    \label{eqn:GasMomen}
\end{equation}
\begin{multline}
    \rho_g c_{p,g} (\frac{\partial \theta_g}{\partial t} + u_i\frac{\partial \theta_g}{\partial x_i})= k_g \frac{\partial^2 \theta_g}{\partial x_i^2} - \sum_{n=1}^N c_{p,g}\theta_{g,n} \frac{\mathrm{d} m_n}{\mathrm{d} t} \delta(\vec{x}-\vec{x}_n) \\
    - \sum_{n=1}^N h_m A_n (\theta_{g,n}-\theta_n) \delta(\vec{x}-\vec{x}_n),
    \label{eqn:GasTemp}
\end{multline}
\begin{equation}
\frac{\partial{c}}{\partial{t}} + u_i \frac{\partial c}{\partial x_i} = D_{vap}\frac{\partial^2 {c}}{\partial{x}_i^2} - \sum_{n=1}^N \left( \frac{\rho_l}{\rho_g} {A}_n \frac{\mathrm{d} {r_n}}{\mathrm{d}{t}} \delta(\vec{x}-\vec{x}_n) \right).
\label{eqn:GasMass}
\end{equation}
For droplets, we apply the spherical point-particle model, and consider the conservation of momentum (Maxey-Riley equation \cite{maxey1983}), energy, and mass as follows:
\begin{equation}
    \frac{\mathrm{d} u_{i,n}}{\mathrm{d} t} = (\beta+1) \frac{\mathrm{D} u_{i,g,n}}{\mathrm{D} t} + (\beta+1) \frac{3 \nu_{air} (u_{i,g,n}-u_{i,n})}{r_n^2}f_d + g\beta \hat{e}_y,
    \label{eqn:DropletMomen}
\end{equation}
\begin{equation}
    \rho_l c_{p,l}V_n\frac{\mathrm{d}\theta_n}{\mathrm{d}t} = \rho_l A_n L\frac{\mathrm{d}r_n}{\mathrm{d}t} + h_m A_n(\theta_{g,n}-\theta_n),
    \label{eqn:DropletTemp}
\end{equation}
\begin{equation}
    \frac{\mathrm{d}r_n}{\mathrm{d}t} = -\frac{D_{vap} Sh_{drop}}{2 r_n}\frac{\rho_g}{\rho_l}\ln \left(\frac{1-c_{fluid}}{1-c_{drop}}\right).
    \label{eqn:DropletDiam}
\end{equation}
The notations we used in equations are as follows: ${u}_i$, ${u}_{i,n}$, and ${u}_{i,g,n}$ are the velocity of gas, droplets, and gas at the location of droplets, respectively. ${\theta}_g$, ${\theta}_n$, and ${\theta}_{g,n}$ are in Kelvin and used to represent the temperature of gas, droplets, and gas at the location of droplets, respectively. $c$ is the vapour mass fraction, ${r_n}$ the droplet radius, ${A}_n$ the surface area of the droplets, ${V}_n$ the volume of the droplets, $m_n$ the mass of the droplets. Also ${p}$ denotes the reduced pressure. $h_m$ is the heat transfer coefficient. $L$ is the latent heat of vaporisation of the liquid. $\rho_l$ and $c_{p,l}$ are the density and specific heat capacity of the droplets assumed to consist of water. $\nu_{air}$ is the kinematic viscosity of air. $k_g$ is the thermal conductivity of gas which relates to the thermal diffusivity of gas $D_g$ by $k_g\equiv D_g \rho_g c_{p,g}$. $\rho_g$ and $c_{p,g}$ are the density and specific heat capacity of gas, Note that $\rho_g c_{p,g}=(\rho_a c_{p,a} + \rho_v c_{p,v})$, with $\rho_a$ and $\rho_v$ being the densities of air and vapour and $c_{p,a}$ and $c_{p,v}$ being the specific heat capacities of air and vapour. $c_{drop}$ and $c_{fluid}$ denote the vapour mass fractions of the droplet and the fluid at the location of the droplet. $\beta$ is a dimensionless measure of the droplet density relative to the fluid density, and is defined as $\beta \equiv 3\rho_g/(\rho_g+2\rho_l)-1$. $f_d$ is the prefactor for the drag corrections defined as $f_d = 1+0.169 \Rey_{\textit{drop}}^{2/3}$ \cite{nguyen2003colloidal}. 

The estimations of $h_m$ and $Sh_{drop}$ for a single spherical droplet are given by the Ranz-Marshall correlations \cite{ranz1952Part2}:
\begin{equation}
    Sh_{drop} = 2 + 0.6 Re_{drop}^{1/2}(\nu_{air}/D_{vap})^{1/3}, \label{eqn:ShRanzMarshall}
\end{equation}
\begin{equation}
    h_m r/(D_gc_{p,g}\rho_g) = 2 + 0.6 Re_{drop}^{1/2}(\nu_{air}/D_{g})^{1/3}, \label{eqn:NuRanzMarshall}
\end{equation}
where we have droplet Reynolds number
\begin{equation}
    Re_{drop} = \frac{|u_{i,g,n}-u_{i,n}| (2r)}{\nu_{air}}.
\end{equation}

To calculate the local relative humidity, the saturated vapour mass fraction $c_{sat,vap}$ is determined by ideal gas law $c_{sat,vap}={P_{sat}}/({\rho_gR\theta_g})$, where R is the specific gas constant of water vapour and $P_{sat}$ is the saturated vapour pressure, determined through Antoine's relation as: 
\begin{equation}
P_\textit{sat}(\theta_g)=10^5\exp\left(11.6834-\frac{3816.44}{226.87+\theta_g-273.15}\right).
\end{equation}

\subsection*{Simulation details}
The computational domain size is $\SI{0.37}{\meter}$ (spanwise length) $\times \SI{0.37}{\meter}$ (height) $\times \SI{1.47}{\meter}$ (streamwise length) with corresponding grid points chosen as $512\times 512\times 2048$. We model the mouth as a circular inlet centred at mid-height of the domain and the cough temporal profile we apply is a gamma distribution as
\begin{equation}
\tilde{U}_\textit{cough}(t) = U_\textit{cough}\alpha t \exp(-\alpha t/4), 
    \label{eqn:ucough}
\end{equation}
where $\alpha = \SI{60.9}{\per \second}$, such that the entire cough process lasts for about \SI{0.6}{\second}. Based on an experimental measurement \cite{duguid1946}, we seeded the respiratory event with droplets with initial diameters ranging between \SI{10}{\micro \meter} and roughly \SI{1000}{\micro \meter}. We employ a similar droplet size distribution as \cite{duguid1946,bourouiba2014} with $5000$ droplets. The droplets are randomly positioned at the inlet area and evenly injected in time with the velocity matching the local inlet velocity. Droplets are assumed not to collide or coalesce since the estimates of the droplet volume fractions give O($10^{-6}$). These droplets also do not set the relative humidity of the puff, which at inflow is assumed to be 100\%.  
}

\showmatmethods{} 

\acknow{The authors thank J.G.M.\,Kuerten for suggesting the idea to study droplet condensation. The authors also thank N.\,Hori for the insightful discussions. This work was funded by the Netherlands Organisation for Health
Research and Development (ZonMW), project number 10430012010022: ``Measuring, understanding \& reducing respiratory droplet spreading'', the ERC Advanced Grant DDD, Number 740479 and by several NWO grants. The funders have no role in study design, data collection and analysis or decision to publish. The simulations were performed on the national e-infrastructure of SURFsara, a subsidiary of SURF cooperation, the collaborative ICT organization for Dutch education and research, the Irish Centre for High-End Computing (ICHEC) and Irene at Tr\`{e}s Grand Centre de calcul du CEA (TGCC) under PRACE project 2019215098.}

\showacknow{} 


\end{document}